# Driven Intrinsic Localized Modes in a Coupled Pendulum Array

R. Basu Thakur, L.Q. English
*Department of Physics & Astronomy, Dickinson College, Carlisle, PA 17013*
A. J. Sievers
*Laboratory of Atomic and Solid State Physics, Cornell University, Ithaca, NY 14853*

**Abstract:** Intrinsic localized modes (ILMs), also called discrete breathers, are directly generated via modulational instability in an array of coupled pendulums. These ILMs can be stabilized over a range of driver frequencies and amplitudes. They are characterized by a π-phase difference between their center and wings. At higher driver frequencies, these ILMs are observed to disintegrate via a pulsating instability, and the mechanism of this breather instability is investigated.

## 1. INTRODUCTION

Intrinsic localized modes (ILMs) are spatially localized, time-periodic excitations in nonlinear lattices analogous to solitons but sensitive to the lattice discreteness due to their extremely sharp localization. The ILM, or discrete breather, has become a very well studied excitation in a wide variety of physical systems and has been firmly established as a conceptual entity on par with the soliton.

Initial studies of ILMs excluded damping or driving - key features from an experimental perspective. Later, analytical and numerical treatments found that the addition of a driver could modify the undriven ILM solutions in nontrivial ways [1-3]. Analytically, perhaps two of the most prominent models for which localized solutions have been demonstrated are the nonlinear Schroedinger equation and the sine-Gordon equation, both of which have proved relevant in the study of mechanical solitons [4]. Both equations have been analyzed with the addition of a driving term [5-7], but both are rooted in the continuum approximation.

In previous experiments on coupled pendulum arrays, a vertical driver excited the pendulums parametrically at around twice the natural resonance frequency [8-10]. In these studies, the particular coupling between pendulums introduced large inter-site anharmonicity. In this paper, we examine a pendulum array with harmonic coupling and horizontal, sinusoidal driving. We observe modulational instability (MI) of the uniform mode which at longer times makes possible the formation of stable ILMs locked to the driver. We characterize these driven ILMs in terms of their spatial profiles for different driving frequencies and amplitudes. Finally, we examine an interesting pulsating instability of the ILM at higher driver frequencies and offer a possible mechanism for this instability [11].

## 2. EXPERIMENTAL DETAILS

The pendulum array is schematically shown in Fig.1. The array consists of 15 pendulums connected by springs which couple them torsionally. Each pendulum is comprised of a brass base piece near the suspension point, a threaded rod screwed vertically into the base, and a brass disk attached to the bottom end of the threaded rod which serves as the weight. The pendulums are suspended on taut piano wire of 0.041" diameter which runs through the base of each pendulum and is held up by the outer frame. Springs of diameter 0.4" slip tightly over sleeves at each end of the base pieces and are then glued onto that sleeve for additional support. No slippage of the spring with respect to the pendulum base was ever observed. The first and last pendulums are coupled only to their one neighbor, the other end resting against a spacer piece to keep it at a fixed distance from the frame; this results in open boundary conditions. Due to the use of a threaded rod, the effective length of the pendulums is adjustable. The weights are calibrated 5 gramm disks.

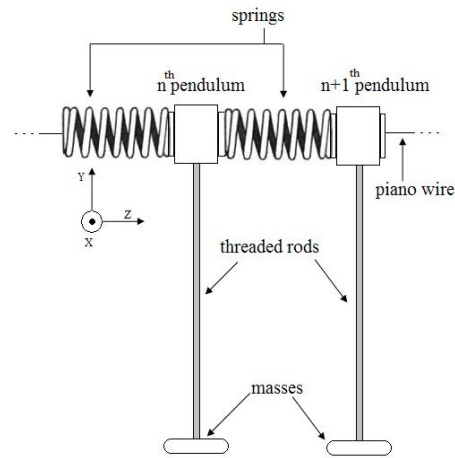

Figure 1 : The experimental system: pendulums connected by torsional springs.

Figure 2 illustrates the larger experimental setup. The pendulum array (i) rests securely on a frame (ii) with low-friction wheels (iii). A high-torque electric motor (90 V, 190 W) is attached to the cart via a crankshaft (iv) and moves it horizontally at a constant but voltage-adjustable frequency. This sinusoidal horizontal motion of the pendulum platform translates into a driving torque on the pendulums. The frequency and amplitude of the driver is independently determined using a Vernier linear motion sensor tracking the pendulum platform. The pendulum motion was captured by an overhead a wide-angle web-cam (v) which was fixed to the frame of the pendulum array (vi). Pendulum coordinates can then be extracted frame-by-frame via video analysis software (Logger Pro, Vernier).

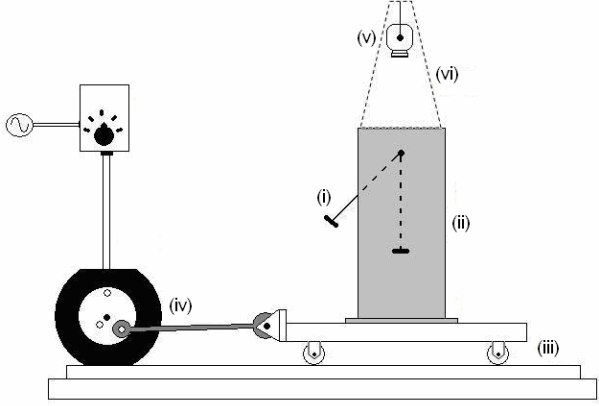

Figure 2 :　　The experimental setup.

Four main torques act on each pendulum: gravitational, spring, damping torques, as well as a centrifugal driving torque by virtue of the non-inertial reference frame. Thus, the equation of motion of the $n^{th}$ pendulum takes the form,

$$\frac{d^2\theta_n}{dt^2} + \omega_0^2 \sin(\theta_n) - \left(\frac{c}{a}\right)^2 [\theta_{n+1} + \theta_{n-1} - 2\theta_n]$$
$$+ \eta \cos(\omega_d t)\cos(\theta_n) + \gamma \frac{d\theta_n}{dt} = 0 \qquad (1)$$

The second term in Eq. (1) is due to the gravitational restoring torque, the third term due to the nearest neighbor spring coupling, the forth incorporates the driver, and the last term accounts for velocity-dependent friction. The spring restoring force was confirmed experimentally to be quite linear up to a ninety degree angle difference, so that the only source of nonlinearity is in the gravitational term. Thus, in contrast to the micromechanical cantilever arrays [12,13], the nonlinearity is strictly on-site. The constants in Eq. (1) can be expressed in terms of measurable properties of our experimental system (see Appendix):

$$\omega_0^2 = \frac{1}{I}\left(mg\frac{L}{2} + MgL\right)$$
$$\eta = \frac{A\omega_d^2}{I}\left(\frac{mL}{2} + ML\right) \qquad (2)$$
$$I = ML^2 + \frac{1}{3}mL^2$$

where $m$ is the mass of the rod, $M$ that of the disk, and $A$ the driving amplitude. The coupling- and friction constants, $c$ and $\gamma$, can be measured directly. The Q-factor of the system is around 120.

Equation (1) is similar to the discrete sine-Gordon equation. The driver's torque, however, introduces a cosine term (instead of the sine for parametric driving [8]), and there is a velocity-dependent damping term.

Equation 1, without the last two terms, results in a linear plane-wave dispersion curve of the form,

$$\omega(k) = \sqrt{\omega_0^2 + 2\left(\frac{c}{a}\right)^2 (1 - \cos(k))} \qquad (3)$$

with $\omega_0$ as given in Eq. (2). Inserting realistic experimental values, we compute the following important frequencies: $\omega_0$ = 6.95 rad/s, $\omega_{ZB}$ = 7.6 rad/s. The driver frequency, $\omega_d$, is in the range of 5.3 to 6.3 rad/s, ensuring that no ILM overtones lie within the linear spectrum.

## 3. RESULTS AND DISCUSSION

Figure 3 illustrates one pathway to stable driven ILMs, namely the modulational instability (MI) of the uniform pendulum mode. The traces depict the positions of the 15 pendulums at the instant the central pendulum reaches its turning point. The first trace in Fig. 3(a) shows the pendulums soon after the driver has been turned on. The driver frequency is set to 99 percent of the linear resonance frequency ($\omega_d$ = 6.85 rad/s). The subsequent traces in Fig. 3(a) show the turning points of the pendulums at later times as they acquire more energy from the driver. In traces 3 and 4, an emerging spatial variation can already be observed. This spatial variation quickly escalates with time, as seen in Fig. 3(b). Traces 5 through 7 show the position of the pendulum chain at three successive periods of the center pendulum. In trace 7 - 11 seconds after the driver is turned on - we see that three pendulums (including the center one) are now out of phase relative to the rest of the chain.

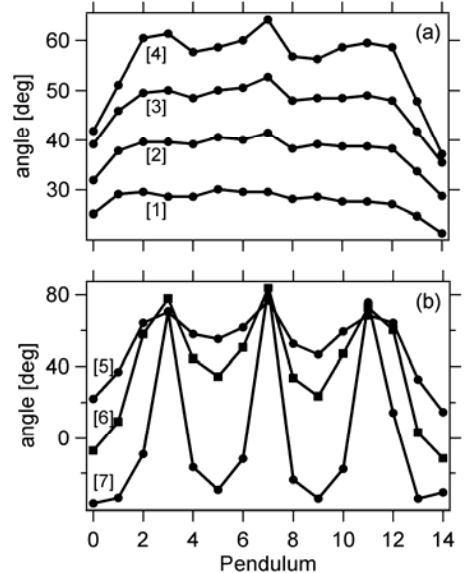

Figure 3 :　　The modulational instability develops as the pendulums are driven first at resonance (upper panel) and then below resonance (lower panel).

The MI essentially amplifies the noise sufficiently for the system to eventually settle into one or more ILMs [3]. It should be emphasized that in order to produce an ILM using the MI, it is essential that the frequency be decreased during the run. We have found experimentally that it is best to start lowering the frequency at the time the spatial variation first becomes noticeable, i.e., somewhere between the traces 3 and 4 of Fig. 3(a). In this run, the frequency was first set to resonance and then lowered within 3 seconds by about 10 percent to $\omega_d = 6.1$ rad/s. In this way, a stable driven ILM was generated in this particular run centered at pendulum 7. The location of the ILM within the chain and even the number of ILMs produced cannot be predicted apriori, and is different from run to run, as expected for a uniform system of discrete nonlinear elements.

Figure 4(a) shows a typical profile of a driven ILM produced via the modulational instability at long times. Evidently, the ILM is localized to about three pendulums. Interestingly, the wings are 180 degrees out of phase with respect to the center of the ILM. Subsequent comparison to the driver phase reveals that the center is out of phase with the driver. In Figure 4(b), another possible stable ILM solution is depicted. Here two ILMs oscillating in phase with one another (but out of phase with the driver) are produced via MI; the ILM peaks are separated by only two pendulums.

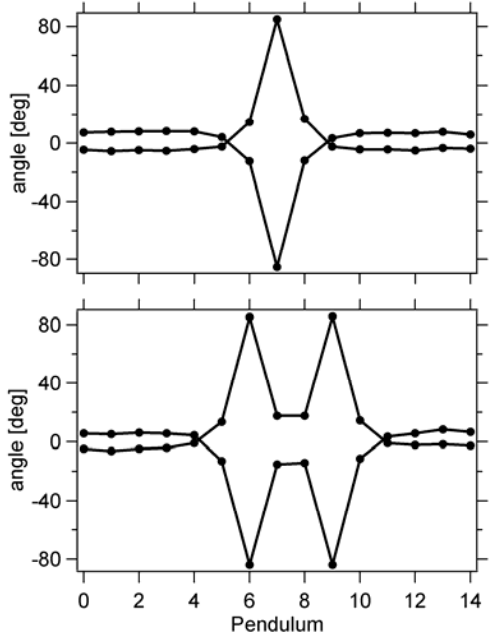

Figure 4 : The long-time profile of driven ILMs. Note that in both cases, the wings perform small out-of-phase oscillations. (a) One ILM at the center of the chain (pendulum 7). (b) Two in-phase ILMs next to one another.

Theoretical studies [1-3] have predicted the existence of ILMs with a 180 degree phase difference between center and wings. This creates what Ref. 1 calls a "phase-domain wall" of just one or two pendulums in width. In systems with soft nonlinearity, the low-amplitude wings are in phase with the driver (as the driver frequency is below the low-amplitude resonance frequency), and the ILM center is out-of-phase. A second type of ILM is also predicted where both center and wings are in-phase with the driver. We have not been able to stabilize this solution experimentally over long times.

Once a stable solution of a driven ILM is obtained experimentally, one can slowly vary the frequency of the driver to see what effect this has on the ILM profile. The results for two driver amplitudes, 7.5 mm (filled circles) and 2.5 mm (open squares), are illustrated in Fig. 5. Note that the x-axis measures the frequency ratio squared. It is evident that stable ILM solutions can be obtained over a limited range of frequencies. When lowering the frequency gradually, the locked ILM adjusts by increasing its amplitude and becoming narrower as it moves away from the plane wave spectrum. Eventually, the ILM simply decouples from the driver and proceeds to damp out as if no driver were present. This happens sooner for lower driver amplitudes, as depicted in the figure.

At the higher limit, the driven ILMs are observed to undergo low-frequency oscillations in amplitude (and width) or pulsations. Above a certain frequency threshold, these pulsations grow in time and eventually destroy the ILM. For low driver amplitudes, this threshold is higher, allowing one to observe broader ILMs. Remarkably, when the results obtained from the two driver amplitudes are superimposed as in Fig. 5, they appear to fall on the same general curve. Perhaps not surprisingly, the data points for the driven ILMs deviate from the solutions to the continuous sine-Gordon equation (dashed line) [4], especially for lower frequencies, i.e., shaper ILMs.

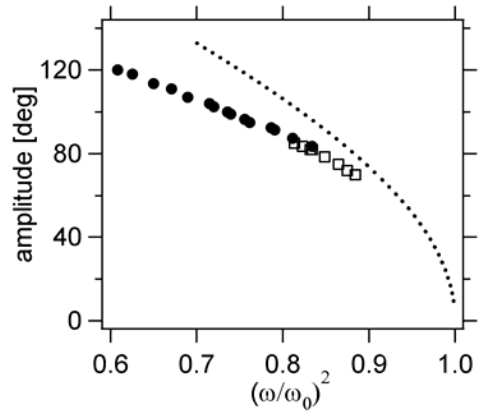

Figure 5 : The amplitude in degrees of the central pendulum of the driven ILM as a function of the driver (and ILM-) frequency. The frequency interval over which ILMs can be produced depends on the driver amplitude. Filled circles represent an amplitude of 7.5mm, open squares 2.5 mm.

When the driver frequency crosses a threshold, unstable ILM pulsations set in. For instance, at $\omega_d = 6.52$ rad/s, corresponding to a frequency ratio squared of 0.91, and amplitude 2.5 mm (see squared dots in Fig. 5), very long-lasting ILM pulsations can be observed. The pulsation period is about 20 times longer than the ILM period, and here the ILM pulsation persists for about 18 minutes gradually growing in modulation amplitude. Just before disintegration, the ILM often hops to a neighboring site temporarily, and soon full-blown chaotic dynamics ensues.

To highlight the nature of this instability, we now change the boundary conditions to fixed. Fig. 6(a) tracks the angle at the turning point, or the maximum angle, for three different pendulums, with the driver set above the frequency threshold for instability. On the bottom, the diamonds represent the center pendulum (pendulum 8), the filled and open circles represent pendulums 4 and 5, respectively. Thus, the upper half of the figure depicts pendulums in the wings of the ILM and the lower half depicts the central pendulum. Note that since none of the markers are at the same time, only amplitude but no phase relationship between the pendulums can be inferred form this figure alone. However, the wings and center tend to be out of phase except around the times of minimum wing amplitude (line labeled (3) in Fig. 6a).

From the diamonds, it is apparent that the amplitude of the ILM is modulated and that this modulation grows with time. The wings undergo the same modulation, but Fig. 6(a) demonstrates that the two modulations are not $\pi$ out of phase, as might be expected for a simple pulsating ILM alternately sharpening and broadening. The situation here is more complex, as illustrated by the profiles shown in Figs. 6(b)-(d).

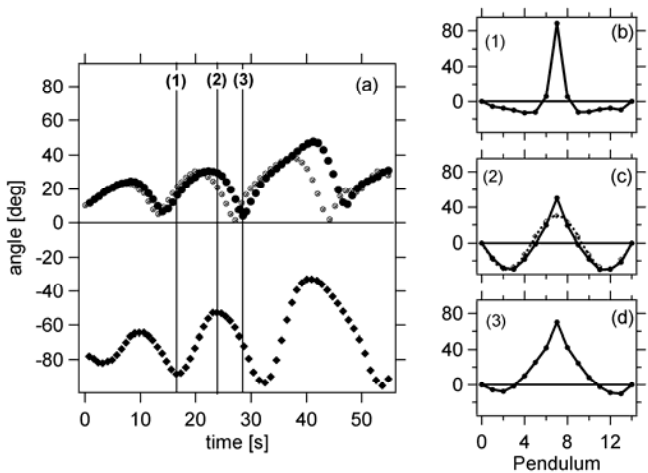

Figure 6 :      (a) ILM instability at higher driver frequencies. The lower trace shows the maximum angle attained by the center pendulum in each period; the two upper traces are for two pendulums in the wings. (b)-(d) The ILM profiles at select times.

Figure 6(b) depicts the position of all pendulums at the time of maximum amplitude of the central pendulum. We see a typical ILM profile with the wings out of phase against the center. Note, however, that in Fig. 6(a) the wings are already gaining amplitude at that time. A while later their amplitude has attained the local maximum while the ILM center is approaching its minimum amplitude, see second vertical line in (a). At this time, the profile is depicted in Fig. 6(c). The solid line depicts the actual pendulum positions, whereas for comparison the dotted line depicts a pure sine wave, the third allowed standing wave in the system with $\lambda_2=2L/3$. We see that with the exception of the central pendulum, the sine-wave fits quite well. Finally, Fig. 6(d) shows the pendulum profile at the time of minimum wing amplitude. Most of the pendulums momentarily move in phase at this time.

The following interpretation of the breather instability mechanism thus emerges (also see Ref. 11). At sufficiently high driver frequencies, the ILM is close enough to the plane-wave band that its presence pulls down the third allowed plane wave of wavelength $\lambda_2$ to the same frequency. At that point –their frequencies matched - energy can be transferred from the ILM to the plane-wave mode. This reduces the ILM amplitude and increases that of the plane wave, as evidenced in Fig. 6(c). Since the ILM amplitude is lowered, the excited plane wave moves back up in frequency. As its phase relationship to the ILM changes, it briefly transfers energy back to the ILM and driver, before the energy transfer stops. Subsequently, the driver can build up the ILM once more.

In short, the nonlinear coupling and decoupling between ILM and third allowed plane-wave mode, as well as both modes' interaction with the driver, give rise to this pulsating breather instability.

## Acknowledgement

This research was supported by an award by Research Corporation. We thank Richard Lindsey for much help in constructing the apparatus.

## APPENDIX

The gravitational restoring torque comes in two parts because we have a rod of non-negligible mass as well as an end-mass. The magnitude of this torque is, $\left(MgL + mg\frac{L}{2}\right)\sin(\theta)$. When dividing by the moment of inertia (which also comes in two parts), we obtain the frequency squared, as in Eq. (2).

The driving torque comes about due to the horizontal motion of the pendulum array by an outside motor. To obtain the expression for the driving torque, we stay within the non-inertial reference frame of the pendulums. In this frame, the fictitious centrifugal force acts on each pendulum, namely $F_c = -ma_{base}$. This force acts on the rod at its center of mass in a horizontal direction, and it thus gives rise to the torque, $\tau_d = A\omega_d^2 \cos(\omega_d t)\left(\frac{mL}{2}\right)\cos(\theta_n)$. For the end-mass, the torque expression is almost identical, except that the force acts at a distance $L$ from the pivot. Thus, we obtain the expression for $\eta$ in Eq. (2).